\documentclass[aps,english,prl,superscriptaddress,onecolumn,preprint,longbibliography]{revtex4-1}
\usepackage[T1]{fontenc}
\usepackage[latin9]{inputenc}
\usepackage{amsmath}
\usepackage{amssymb}
\usepackage{graphicx}
\usepackage{babel}
\begin{document}
\preprint{Draft \#1}
\title{Controlling cell motion and microscale flow with polarized light fields }
\author{Siyuan Yang}
\affiliation{School of Physics and Astronomy and Institute of Natural Sciences,
Shanghai Jiao Tong University, Shanghai 200240, China}
\author{Mingji Huang}
\affiliation{School of Physics and Astronomy and Institute of Natural Sciences,
Shanghai Jiao Tong University, Shanghai 200240, China}
\author{Yongfeng Zhao}
\affiliation{School of Physics and Astronomy and Institute of Natural Sciences,
Shanghai Jiao Tong University, Shanghai 200240, China}
\author{H. P. Zhang}
\email{hepeng\_zhang@sjtu.edu.cn}

\affiliation{School of Physics and Astronomy and Institute of Natural Sciences,
Shanghai Jiao Tong University, Shanghai 200240, China}
\affiliation{Collaborative Innovation Center of Advanced Microstructures, Nanjing 210093,
China}
\date{\today}
\begin{abstract}
We investigate how light polarization affects the motion of photo-responsive
algae, \textit{Euglena gracilis}. In a uniformly polarized field,
cells swim approximately perpendicular to the polarization direction
and form a nematic state with zero mean velocity. When light polarization
varies spatially, cell motion is modulated by local polarization.
In such light fields, cells exhibit complex spatial distribution and
motion patterns which are controlled by topological properties of
the underlying fields; we further show that ordered cell swimming
can generate directed transporting fluid flow. Experimental results
are quantitatively reproduced by an active Brownian particle model
in which particle motion direction is nematically coupled to local
light polarization. 
\end{abstract}
\maketitle
Natural microswimmers, such as bacteria and algae, can achieve autonomous
motion by converting locally stored energy into mechanical work \citep{Lauga2009a,Ramaswamy2010,Poon2013,Aranson2013,Wang2013c,Sanchez2015,Elgeti2015,RevModPhys.88.045006,Lavrentovich2016,Zottl2016,Patteson2016,Zhang2017b,C7CS00087A,Liebchen2018a,Gompper2020}.
Such cellular motility is not only an essential aspect of life but
also an inspirational source to develop artificial microswimmers,
which propel themselves through self-generated fields of temperature,
chemical concentration, or electric potential \citep{Lauga2009a,Poon2013,Aranson2013,Wang2013c,Sanchez2015,Elgeti2015,Zhang2017b,C7CS00087A}. Both natural and artificial microswimmers have been used in a wide variety of applications \citep{Wang2012d,Gao2014,Li2017d,Alapan2019}.

To properly function in a fluctuating heterogeneous environment, microswimmers
need to adjust their motility in response to external stimuli \citep{Menzel2015,Stark2016,You2018,Klumpp2019}.
For example, intensity and direction of ambient light can induce a
variety of motility responses in photosynthetic microorganisms \citep{Mikolajczyk1990,Jekely2009,Drescher2010a,Barsanti2012,Kane2013,Garcia2013,Giometto2015,Bennett2015,Chau2017,Hader2017,10.1371/journal.pone.0172813,Arrieta2017,Tsang2018,Arrieta2019,Choudhary2019}
and artificial microwimmers \citep{C7CS00516D,Dong2018,Wang2018a,Aubret2018,Singh2018,Zhan2019,Lavergne2019};
these responses have been frequently used to control microswimmer
motion \citep{Arlt2018,Tsang2018,Arrieta2017,Dervaux2017,10.1371/journal.pone.0168114,Stenhammar2016,Palacci2013,Frangipane2018,10.1371/journal.pone.0172813,Lozano,Giometto2015,Geiseler2016,Barsanti2012,Lavergne2019}.
Besides intensity and direction, light \textit{polarization} can also
affect microswimmer motility and lead to polarotaxis: \textit{Euglena
gracilis} cells align their motion direction perpendicular to the
light polarization, possibly to maximize the light absorption \citep{CREUTZ1976,Hader1987};
artificial microswimmers consisting of two dichroic nanomotors move
in the polarization direction \citep{Zhan2019}. These previous experiments
have focused on uniform light fields \citep{CREUTZ1976,Hader1987,Zhan2019}.
The possibility to use complex polarization patterns to control polarotactic
microswimmers has not been explored.

In this letter, we investigate \textit{Euglena gracilis} cell motion
in various polarized light fields in a quantitative and systematic
fashion. Our experiments show that while spatially uniform polarization
aligns cells into a global nematic state with no net motion, spatially
varying fields can induce both local nematic order and mean cell motion. Further, we show that ordered cell swimming motion generates
fluid flow that can transport passive tracers. Using the experimental
data of individual cells, we construct a model to describe the
influence of local light polarization on cell orientation dynamics
and quantitatively reproduce all experimental observations. 

\textit{Experiments} - \textit{Euglena gracilis} are unicellular flagellated
microorganisms with a rod-shaped body of a length $\sim$ 50 $\mu$m
and a width $\sim$ 5 $\mu$m. As shown in Fig. 1(a) and Movie S1
in the Supplemental Material \citep{SI}, cells swim at a mean speed
$\sim$ 60 $\mu$m/s (with a standard deviation of $10\text{ }\mu$m/s.),
while rolling around their long axis at a frequency of 1-2 Hz \citep{Rossi2017}.
A photoreceptor on \textit{Euglena} cell surface, marked as a red
dot in Fig. 1(b), senses surrounding light and generate signals to
modulate flagellar beating pattern \citep{Hill2000,Hader2017}. 

In our experiments\textit{, Euglena }culture is sealed in a disk-shaped
chamber ($\sim150\text{ }\mu$m in thickness and 24 mm in diameter),
which is placed in an illuminating light path, as shown in Fig. S1 \citep{SI}. A collimated blue light
beam is used to excite cell
photo-responses; the default light intensity is 100 $\mu\text{W/cm}^{2}$.
Various polarized optical fields can be generated by using different
birefringent liquid crystal plates and by changing relative angles
between optical elements \citep{Delaney2017}. Cell motion is recorded
by a camera mounted on a Macro-lens. Default system cell density ( $\rho_{0}=$ 8 cells/mm$^{2}$ ) is sufficiently low that we can use
a standard particle tracking algorithm \citep{Zhang2010a} to measure
position, orientation, and velocity of cells. The current work mainly
focuses on steady state dynamics that is invariant over time.

\begin{figure}
\centering{}\includegraphics[width=8.5cm]{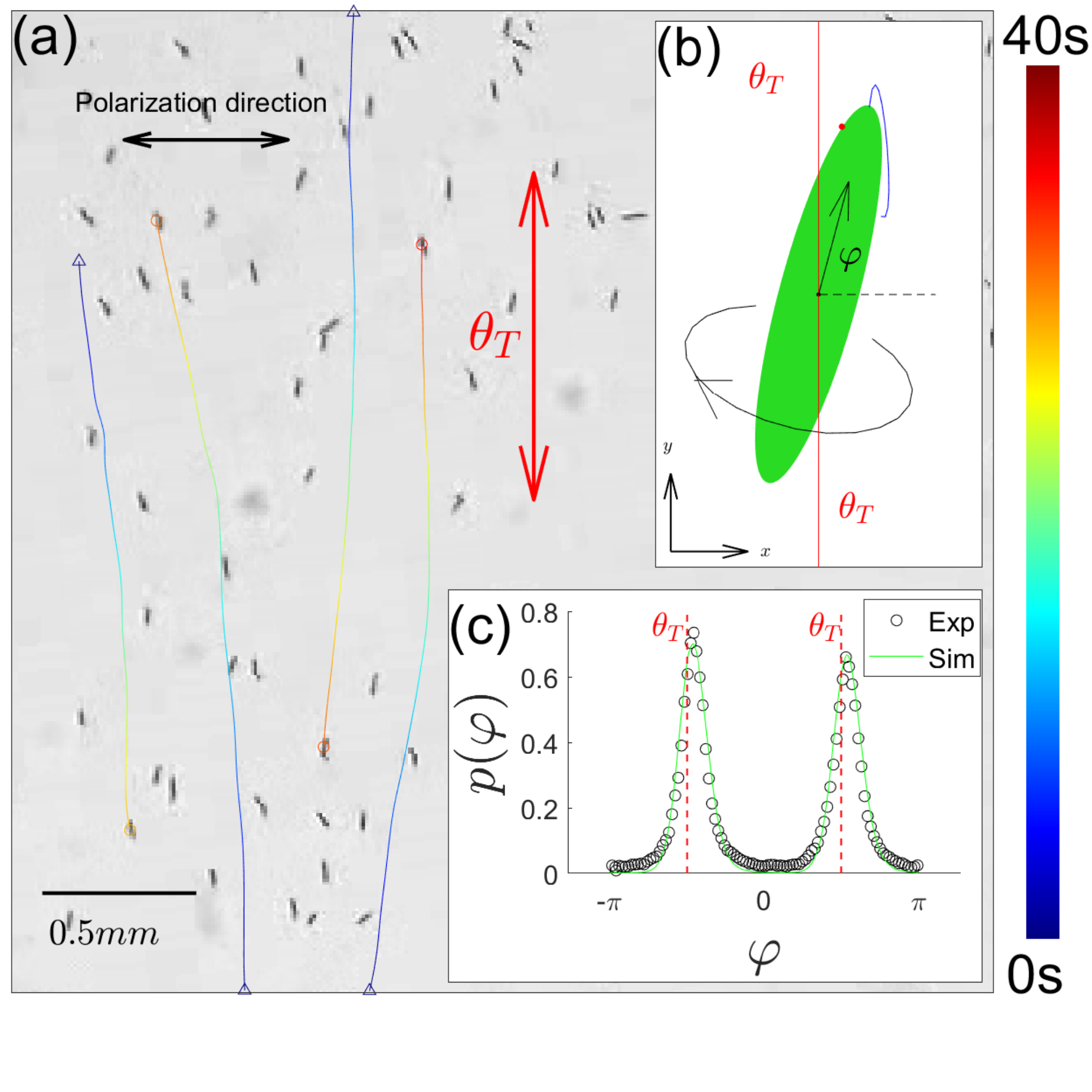}\protect\protect\caption{Cell motion in a uniformly polarized light field. (a) Cell trajectories
(color-coded by time) plotted on an experimental snapshot. Light polarization
is horizontal and cells tend to swim vertically in the targeted direction
$\theta_{T}$. (b) shows a schematic for a cell (with a red eye-spot
and a flagellum) which moves at a $\varphi$ direction; a circular
arrow indicates body rolling motion. (c) Probability distribution
of cell motion direction $\varphi$. }
\end{figure}

\textit{Uniformly polarized light field} -\textit{ Euglena} photoreceptor
contains dichroically oriented chromoproteins which lead to polarization-dependent
photo responses \citep{BOUND1967,CREUTZ1976,Hader1987,Hader2017}.
As shown in Fig. 1(a), cells in a horizontally polarized field tend
to orient and swim perpendicularly to the polarization \citep{CREUTZ1976};
we denote such a targeted direction for cells as $\theta_{T}$. Quantitatively,
we measure the $j$th cell's location $\vec{r}_{j}\left(t\right)$,
velocity $\vec{v}_{j}$, and velocity angle $\varphi_{j}$, cf. Fig.
1(b). Over a square window (1.2 mm$^{2}$), we define mean cell velocity
as $\vec{v}=\left\langle \vec{v}_{j}\right\rangle $, where average
$\left\langle \cdot\right\rangle $ runs over all cells in the region
during the measurement time; nematic order parameter and orientation
angle are defined as $u=\text{\ensuremath{\left|\left\langle \exp\left(\text{i}\left(2\varphi_{j}\right)\right)\right\rangle \right|}}$
and $\phi_{u}=\frac{1}{2}\text{Arg}\left(\left\langle \exp\left(\text{i}\left(2\varphi_{j}\right)\right)\right\rangle \right)$,
where Arg denotes the phase angle of a complex number. In uniform
fields, cells are homogeneously distributed over space and form a
global nematic state with a vanishing mean cell velocity: $u\approx0.75$
and $\vec{v}\approx0$.

\begin{figure}
\centering{}\includegraphics[width=8.5cm]{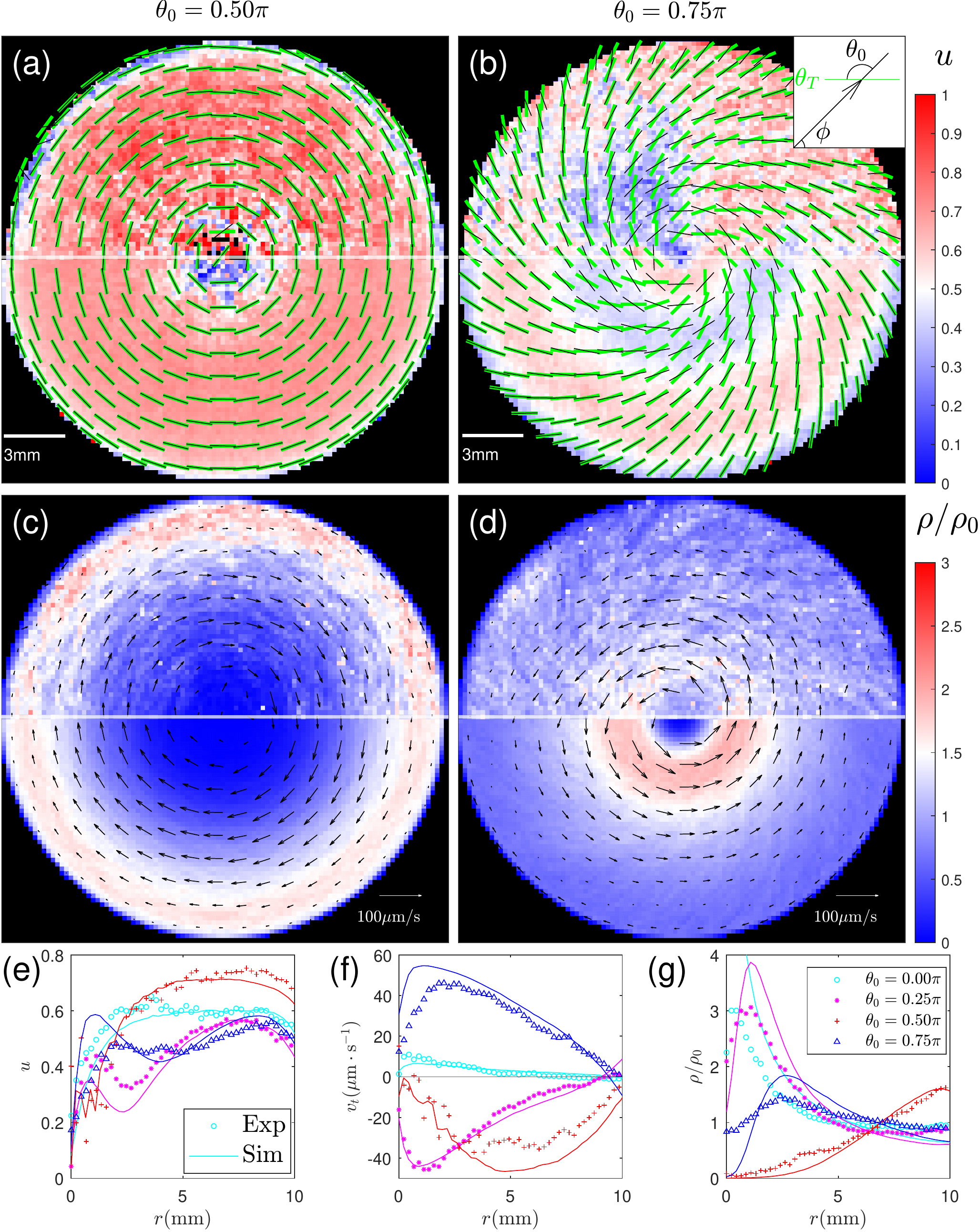}\protect\protect\caption{Orientation, velocity, and cell density in axisymmetric light fields
containing a $k=+1$ defect with $\theta_{0}=\pi/2$ (a,c) and $\theta_{0}=3\pi/4$
(b,d). In (a-b), targeted direction $\theta_{T}$ and mean cell motion
direction $\phi_{u}$ are shown by green and black lines, respectively,
on nematic order parameter $u$ (in color). In (c-d), mean cell velocity
$\vec{v}$ is plotted on mean density (in color). In (a-d), top and
bottom halves (separated by a white line) are experimental and numerical
results, respectively. The inset of (b) defines three angles (see
text). (e-g) Radial profiles of nematic order parameter $u$, tangential
velocity $v_{t}=\vec{v}\cdot\hat{\phi}$, and cell density $\rho$
for four fields.}
\end{figure}

\textit{Axisymmetric light field} - We next investigate cell motion
in light fields with spatially varying polarization. In our experiments,
the targeted direction field $\theta_{T}\left(\vec{r}\right)$ is
designed to have the form of $\theta_{T}\left(\vec{r}\right)=k\phi\left(\vec{r}\right)+\theta_{0}$,
where $k$ is a winding number, $\phi=\tan^{-1}\left(y/x\right)$
is the polar angle, and $\theta_{0}$ is a spiral angle (cf. inset of
Fig. 2(b)). When $k=1$, $\theta_{T}\left(\vec{r}\right)$ field is
axisymmetric as shown by short green lines in Fig. 2 (a-b) and $\theta_{0}$
controls the ratio between bend and splay strength.

Cell motion in axisymmetric fields can be seen in Movies S2-S5 \citep{SI}. Quantitatively, mean nematic
order parameter, cell velocity, and cell density are plotted in Fig.
2 and Fig. S3 \citep{SI}. As shown in
Fig. 2(e), nematic order parameter $u$ increases from the defect
center to the exterior of the illuminated region, where spatial gradients
of $\theta_{T}\left(\vec{r}\right)$ are small and cells closely follow
$\theta_{T}\left(\vec{r}\right)$. Cells in pure bend ($\theta_{0}=\pi/2$)
and mixed ($\theta_{0}=3\pi/4$) light fields also exhibit mean velocity;
peak value in radial profiles in Fig. 2(f) is about $50\text{ }\mu$m/s.
Spatial distributions of cells depend on $\theta_{0}$: while cells
aggregate at the exterior boundary for $\theta_{0}=\pi/2$, Fig. 2(g)
shows a relatively flat distribution with a small peak at $r=2.6$
mm for $\theta_{0}=3\pi/4$ and cell aggregation near the defect center
for other two $\theta_{0}$ conditions. We also systematically vary
light intensity and system cell density; qualitatively similar results are shown in Figs. S4-S5 and Movie S7 in the
Supplemental Material \citep{SI}.  

\begin{figure}
\centering{}\includegraphics[width=8.5cm]{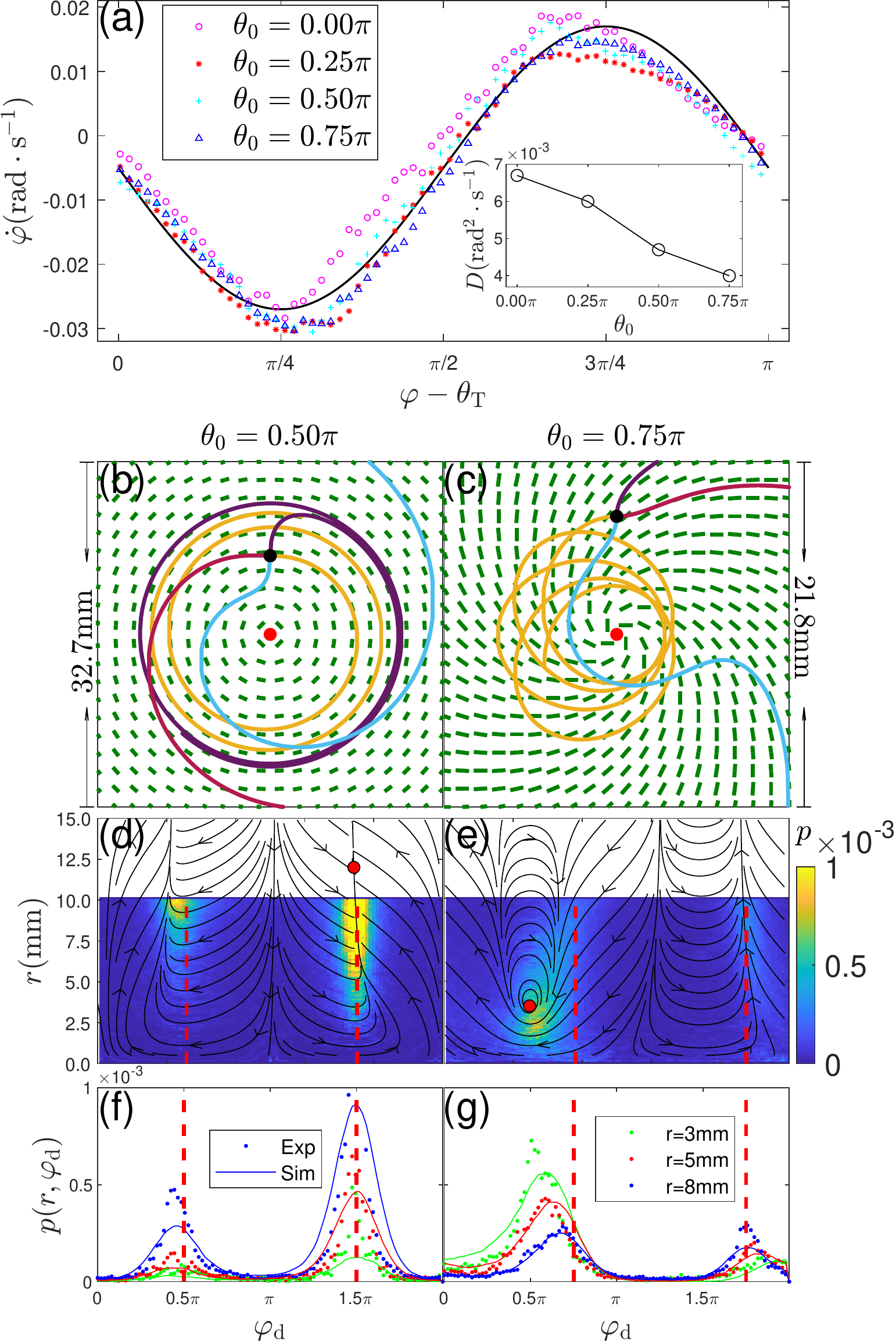}\protect\protect\caption{(a) Mean angular velocity $\dot{\varphi}$ versus the angular deviation
$\varphi-\theta_{T}$ in in axisymmetric light fields. Inset shows
effective diffusivity $D$ measured in different fields $\theta_{0}$.
(b-g) Deterministic trajectory and probability distribution in axisymmetric
fields with $\theta_{0}=\pi/2$ (b, d, f) and $\theta_{0}=3\pi/4$
(c, e, g). (b-c) Cell trajectories from the deterministic model plotted
on the targeted field. See Fig. S8 \citep{SI}
for more trajectories. (d-e) Experimentally measured probability $p\left(r,\varphi_{d}\right)$
(color) and computed phase trajectories (black lines). Stable and
neutrally stable fixed points are colored in red. Fixed points in
(d) is outside of the experimentally measured range ($r<10.8\text{ }\mu$m).
(f-g) Profiles of $p\left(r,\varphi_{d}\right)$ at three radii. Dashed
lines in (d-g) mark targeted direction $\theta_{T}$. }
\end{figure}

\textit{Deterministic model }- Fig. 1 and Fig. 2 show that cells tend
to align their motion direction $\varphi$ towards the local targeted
direction $\theta_{T}\left(\vec{r}\right)$. To quantify this nematic
alignment interaction, we extract the time derivative of motion direction
$\dot{\varphi}_{j}$ from cell trajectories and find that $\dot{\varphi}_{j}$
is a function of the angular deviation $\varphi_{j}-\theta_{T}\left(\vec{r}_{j}\right)$.
We average the dependence function over all cells in a given experiment.
Mean $\dot{\varphi}$ in Fig. 3(a) can be adequately described by
the following equation: 
\begin{equation}
\dot{\varphi}=-A\sin\left(2\left(\varphi-\theta_{T}\right)\right)+C.
\end{equation}
Fitting data in Fig. 3(a) leads to a nematic interaction strength
$A=0.022\text{ rad/s}$ \citep{Li2019} and a constant angular velocity
$C=-0.005\text{ rad/s}$ for default light intensity; parameter $A$ increases with light intensity, and $C$ shows
a weak dependence, as shown in Fig. S4(e)
\citep{SI}. Small negative $C$ value indicates that cells have a
weak preference to swim clockwise; such chirality has been reported
before \citep{Tsang2018} and is likely caused by the symmetry breaking
from handedness of cell body rolling and directionality of the illuminating
light, cf. Fig. S1 \citep{SI}. This
weak chirality explains the non-zero mean cell velocity in an achiral
light field in Fig. 2(c) ($\theta_{0}=\pi/2$). To describe cell
translational motion in our model, we assume all cells have the same
speed $v_{\circ}=60\text{ }\mu$m/s and update cell's position with
a velocity 
\begin{equation}
\dot{\vec{r}}=v_{\circ}\left(\cos\varphi\hat{x}+\sin\varphi\hat{y}\right).
\end{equation}

In axisymmetric fields, particle dynamics from Eqs. (1-2)
can be described by two variables: the radial coordinate $r$ and
the angular deviation from the local polar angle $\varphi_{d}=\varphi-\phi$.
We solve the governing equations for these quantities (cf. the Supplemental
Material \citep{SI}) and compute particle trajectories in $\left(r,\varphi_{d}\right)$
phase plane, as shown dark lines in Fig. 3(d-e). Fixed point in the
phase plane is identified at $r^{*}=\left|\frac{v_{\circ}}{C-A\sin2\theta_{0}}\right|$
and $\varphi_{d}^{*}=\frac{\pi}{2}$ (if $C>A\sin2\theta_{0}$) or
$\varphi_{d}^{*}=-\frac{\pi}{2}$ (if $C<A\sin2\theta_{0}$); it is stable if $\cos2\theta_{0}<0$, neutrally stable if $\cos2\theta_{0}=0$,
and unstable if $\cos2\theta_{0}>0$. At stable and neutrally stable
fixed points, particle moves along circular trajectories, cf. the
violet trajectory in Fig. 3(b). Around neutrally stable fixed points,
there is a family of closed trajectories in $\left(r,\varphi_{d}\right)$
phase plane; in real space, such trajectories appear to be processing
ellipses around the defect center, cf. yellow trajectories in Fig.
3(c) and Fig. S8(c) \citep{SI}.

\textit{Langevin model }- Cell motion contains inherent noises, which
may arise from flagellum dynamics or cell-cell interactions. To account
for this stochasticity, we add a rotational noise term $\sqrt{2D}\xi\left(t\right)$
to Eq. (1), which becomes Eq. (S1) \citep{SI};
$\xi\left(t\right)$ represents Gaussian white noise with zero-mean
$\left\langle \xi\left(t\right)\xi\left(0\right)\right\rangle =\delta\left(t\right)$
and $D$ is an effective rotational diffusivity. With this noise term,
Eq. (S1) and Eq. (2) constitute a Langevin model of an active Brownian
particle whose orientation is locally modulated by the light polarization,
i.e. $\theta_{T}$. The corresponding Fokker-Planck equation can be
written down for the steady-state probability density, $p\left(\vec{r},\varphi\right)$, of finding a
particle at a state $\left(\vec{r},\varphi\right)$.
For uniformly polarized field, the probability distribution $p\left(\varphi\right)$
can be analytically solved and fitted to data in Fig. 1(c), yielding an estimation of $D/A=0.17 \text{\text{ }rad}$ for this experiment.

We then consider axisymmetric fields. Probability density $p\left(r,\varphi_{d}\right)$
is experimentally measured and Fig. 3 (d-e) show high value around
stable/neutrally stable fixed points. This highlights the importance
of fixed points: their radial positions determine cell distributions
in Fig. 2(c-d) and they appear at either $\varphi_{d}^{*}=+\frac{\pi}{2}$
or $\varphi_{d}^{*}=-\frac{\pi}{2}$, which breaks the chiral symmetry
and leads to a non-zero mean velocity. $p\left(r,\varphi_{d}\right)$
measured in two other cases of $\theta_{0}$ are shown in Fig. S3
\citep{SI}. To quantitatively reproduce
measured $p\left(r,\varphi_{d}\right)$, we numerically integrate
the Langevin model: parameters $A$ and $C$ values extracted from
Fig. 3(a) are used and the effective angular diffusivity $D$ is tuned
to fit experimental measurements, see inset of Fig. 3(a). Our numerical
results agree well with experiments for probability density profiles
in Fig. 3 (f-g) and for radial profiles in Fig. 2 (e-g). 

\begin{figure}
\centering{}\includegraphics[width=8.5cm]{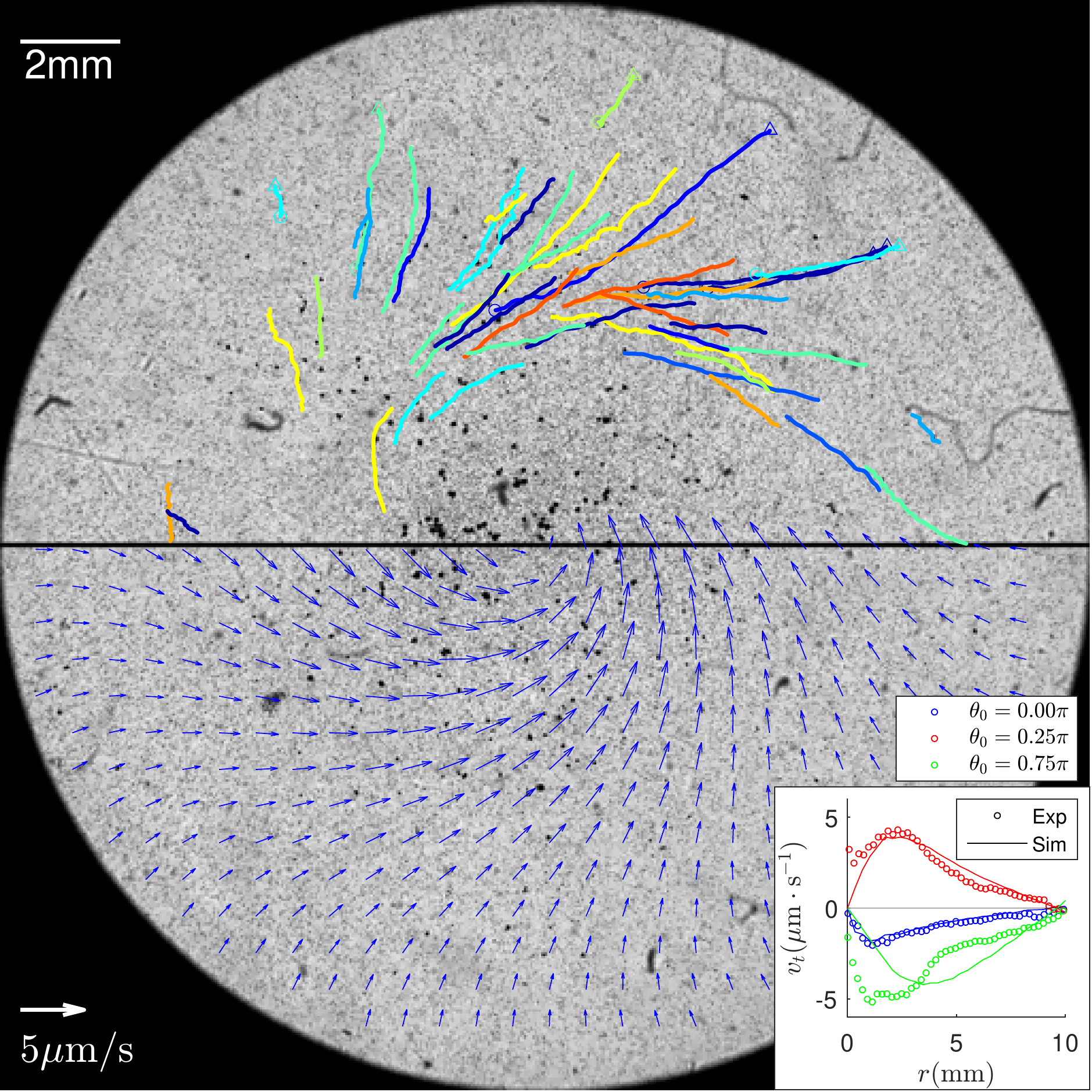}\protect\protect\caption{Trajectories of passive tracers (top panel, from experiments) and
flow field (bottom panel, from the dipole model) driven by \textit{Euglena}
in a light field with $k=+1$ and $\theta_{0}=\pi/4$. An experimental
snapshot is shown in the background. The inset shows radial profiles
of tracers tangential velocities in three axisymmetric ($k=1$) light
fields.}
\end{figure}

\textit{Transport of passive particles }- Ordered swimming of \textit{Euglena}
cells in Fig. 2 can collectively generate fluid flow \citep{Mathijssen2018},
which we use hollow glass spheres (50 $\mu$m) on an air-liquid interface
to visualize. Tracer trajectories from an experiment are shown in
the top half of Fig. 4 and particles spiral counter-clock-wisely
towards the center with a peak speed about 5 $\mu$m/s. To compute
the generated flow, we represent swimming cells as force-dipoles
\citep{Ogawa2017,Bardfalvy2020}: a dipole in a state $\left(\vec{r},\varphi\right)$
generate flow velocity $\vec{w}\left(\vec{r}_{s};\vec{r},\varphi\right)$
(including contributions from a force-dipole \citep{Ogawa2017} and
its image \citep{HappelBook, Mathijssen2015}) at a location on the surface $\vec{r}_{s}$.
Then, for a given light field, the Langevin model is used to simulate the motion of $N$ cells and 
to find the probability distribution of cells $p\left(\vec{r},\varphi\right)$. Finally, we compute the total flow as:
$\vec{W}\left(\vec{r}_{s}\right)=N\int p\left(\vec{r},\varphi\right)\vec{w}\left(\vec{r}_{s};\vec{r},\varphi\right)\text{d}\text{\ensuremath{\vec{r}}d\ensuremath{\varphi}}$,
see Sec. II(F) in the Supplemental Material \citep{SI} for details. This
approach generates flow fields (cf. bottom half and inset of Fig. 4) that are
consistent with measured tracer velocities, see also Fig. S6 \citep{SI}.

\begin{figure}
\centering{}\includegraphics[width=8.5cm]{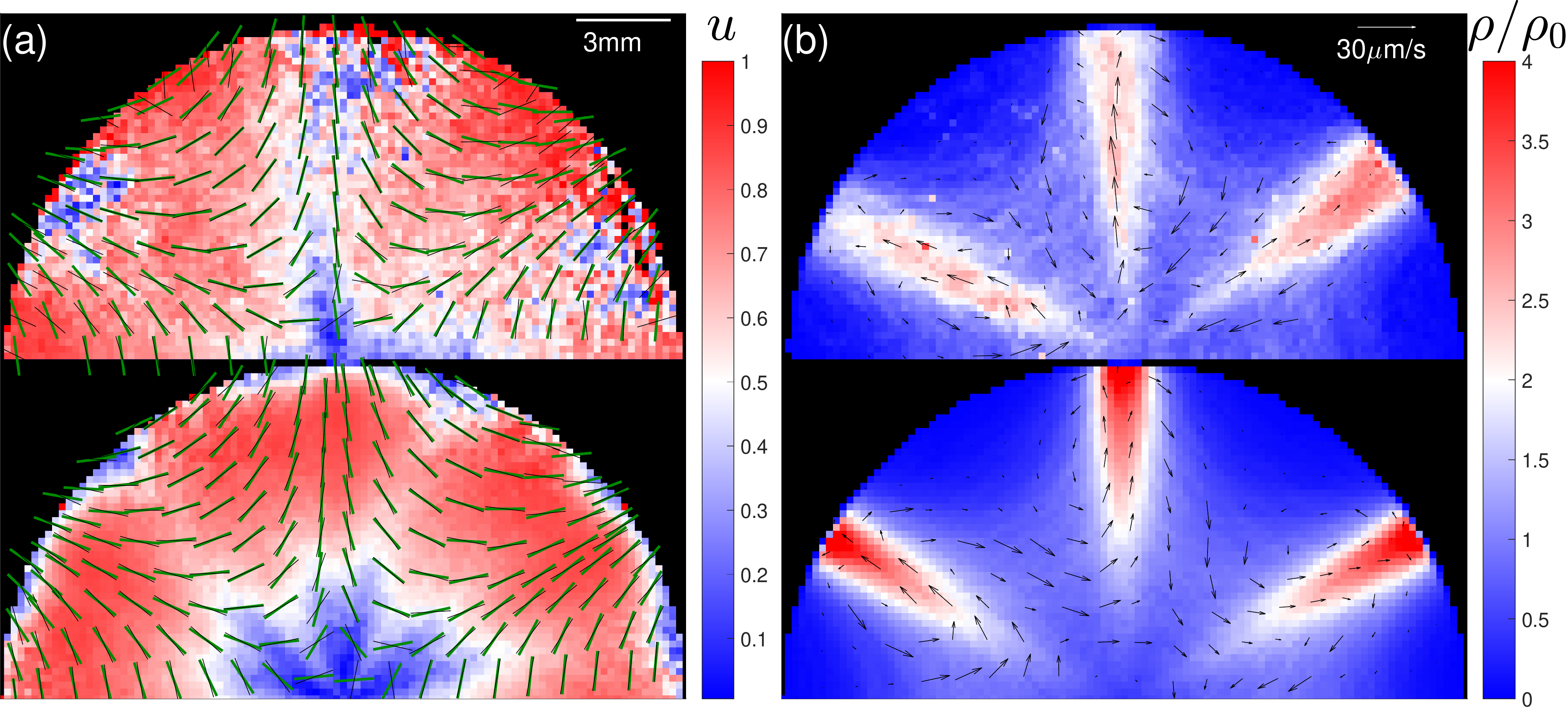}\protect\protect\caption{Orientation (a) and velocity/density (b) in a light fields containing
a $k=-2$ defect with $\theta_{0}=\pi/2$. In (a), targeted direction
$\theta_{T}$ and mean cell motion direction $\phi_{u}$ are shown
by green and black lines, respectively, on nematic order parameter
$u$ (in color). In (b), mean cell velocity $\vec{v}$ is plotted
on mean density (in color). Top and bottom panels are experimental
and numerical results, respectively.}
\end{figure}

\textit{Discussion }- Our setup can also
generate nonaxisymmetric light fields with integer winding numbers.
Fig. 5 shows that cells in a \textit{$k=-2$ }field form dense and
outgoing bands in regions where $\theta_{T}$ is close to be radial;
these observations can be explained by stable radial particles trajectories
in Fig. S9 (also Movie S6) \citep{SI}.
The Langevin model is used to investigate light fields with half-integer
defects and multiple defects \citep{Rosales-Guzman2018}; results
of cell dynamics and transporting flow in Figs. S12 and S13 \citep{SI}
demonstrate that our idea of local orientation modulation can be used
as a versatile and modular method for system control.

Local orientation modulation has been previously implemented by embedding
rod-shaped bacteria in nematic liquid crystal with patterned molecular
orientation \citep{Trivedi2015,Peng2016,Aranson2018,Turiv2020,Koizumi2020}.
In this bio-composite system, while cell orientation is \textit{physically}
constrained by aligned molecules, bacteria swimming can in return
disrupt the molecular order; this strong feedback weakens the controlling
ability of the imposed pattern and leads to highly complex dynamics
\citep{Trivedi2015,Peng2016,Aranson2018,Turiv2020,Koizumi2020}. By
contrast, our method relies on \textit{biological responses}, instead
of physical interactions, to achieve orientation control, and\textit{
Euglena} motion has no effect on the underlying light field. Such
a one-way interaction leads to a much simpler system and may help
us to achieve more accurate control. Furthermore, our method works
on cells in their natural environment and requires no elaborate sample
preparation. This factor and the spatio-temporal tunability of light
fields \citep{Rosales-Guzman2018} make our method flexible and easy
to use.

Sinusoidal term in Eq. (1) is the simplest harmonic for nematic alignment.
The same term has been observed in dichroic nano-particle systems
\citep{Tong2010,Zhan2019} and is related to the angular dependence
of dichroic light absorption. These nano-particle systems usually
require very strong ($\sim$ $\text{W/cm}^{2}$-$\text{MW/cm}^{2}$
) light stimulus to operate. By contrast, biological response in \textit{Euglena}
greatly amplifies the light signal and functions in the range of 100
$\mu\text{W/cm}^{2}$; this high sensitivity significantly
reduces the complexity to construct a controlling light field.

\textit{Conclusion }- To summarize, we have experimentally demonstrated
that \textit{Euglena} motion direction is strongly affected by the
local light polarization and that cell dynamics in spatially varying
polarization fields is controlled by topological properties and light
intensity of the underlying fields. Our experiments also showed that
ordered cell swimming, controlled by the polarization field, can generate
directed transporting fluid flow. Experimental results have been quantitatively
reproduced by an active Brownian particle model in which particle
motion direction is nematically coupled to the local light polarization;
fixed points and closed trajectories in the model have strong impacts
on system properties. These results suggest that local orientation
modulation, via polarized light or other means, can be used as a general
method to control active matter and micro-scale transporting flow.

\begin{acknowledgments}
\textit{Acknowledgments - }We acknowledge financial support from National
Natural Science Foundation of China (Grants No. 11774222 and No. 11422427)
and from the Program for Professor of Special Appointment at Shanghai
Institutions of Higher Learning (Grant No. GZ2016004). We thank Hugues
Chat\'e and Masaki Sano for useful discussions and the Student Innovation
Center at Shanghai Jiao Tong University for support.
\end{acknowledgments}

\bigskip{}

\bibliographystyle{apsrev4-1}
\end{document}